\documentclass[prl,twocolumn,superscriptaddress,noshowpacs,floatfix]{revtex4}
\usepackage{epsfig}
\usepackage{dcolumn}
\usepackage{color}

\begin{document}

\title{Fusion reactions with the one-neutron halo nucleus $^{15}$C}
  
\author{M.~Alcorta}
\affiliation {Physics Division, Argonne National Laboratory, Argonne, IL 60439, USA}
\author{K.~E.~Rehm}
\affiliation {Physics Division, Argonne National Laboratory, Argonne, IL 60439, USA}
\author{B.~B.~Back}
\affiliation {Physics Division, Argonne National Laboratory, Argonne, IL 60439, USA}
\author{S.~Bedoor}
\affiliation {Western Michigan University, Kalamazoo, MI 49008, USA}
\author{P.~F.~Bertone}
\affiliation {Physics Division, Argonne National Laboratory, Argonne, IL 60439, USA}
\author{C.~M.~Deibel}
\affiliation {Physics Division, Argonne National Laboratory, Argonne, IL 60439, USA}
\affiliation{Joint Institute for Nuclear Astrophysics, Michigan State University, East Lansing, MI 48824, USA}
\author{B.~DiGiovine}
\affiliation {Physics Division, Argonne National Laboratory, Argonne, IL 60439, USA}
\author{H.~Esbensen}
\affiliation {Physics Division, Argonne National Laboratory, Argonne, IL 60439, USA}
\author{J.~P.~Greene}
\affiliation {Physics Division, Argonne National Laboratory, Argonne, IL 60439, USA}
\author{C.~R.~Hoffmann}
\affiliation {Physics Division, Argonne National Laboratory, Argonne, IL 60439, USA}
\author{C.~L.~Jiang}
\affiliation {Physics Division, Argonne National Laboratory, Argonne, IL 60439, USA}
\author{J.~C.~Lighthall}
\affiliation {Physics Division, Argonne National Laboratory, Argonne, IL 60439, USA}
\affiliation {Western Michigan University, Kalamazoo, MI 49008, USA}
\author{S.~T.~Marley}
\affiliation {Physics Division, Argonne National Laboratory, Argonne, IL 60439, USA}
\affiliation {Western Michigan University, Kalamazoo, MI 49008, USA}
\author{R.~C.~Pardo}
\affiliation {Physics Division, Argonne National Laboratory, Argonne, IL 60439, USA}
\author{M.~Paul}
\affiliation {Racah Institute of Physics, Hebrew University, Jerusalem, 91904, Israel}
\author{A.~M.~Rogers}
\affiliation {Physics Division, Argonne National Laboratory, Argonne, IL 60439, USA}
\author{C.~Ugalde}
\affiliation {Physics Division, Argonne National Laboratory, Argonne, IL 60439, USA}
\affiliation{Department of Astronomy and Astrophysics, University of Chicago, Chicago, IL 60637 USA}
\affiliation{Joint Institute for Nuclear Astrophysics, Chicago, IL 60637 USA}
\author{A.~H.~Wuosmaa}
\affiliation {Western Michigan University, Kalamazoo, MI 49008, USA}

\date{\today}
%\maketitle

\begin{abstract}
The structure of $^{15}$C, with an s$_{1/2}$ neutron weakly bound to a closed-neutron shell nucleus $^{14}$C, makes it a prime candidate for a one-neutron halo nucleus. We have for the first time studied the cross section for the fusion-fission reaction $^{15}$C + $^{232}$Th at energies in the vicinity of the Coulomb barrier and compared it to the yield of the neighboring $^{14}$C + $^{232}$Th system measured in the same experiment. At sub-barrier energies, an enhancement of the fusion yield by factors of 2-5 was observed for $^{15}$C, while the cross sections for $^{14}$C match the trends measured for $^{12,13}$C.  
\end{abstract}
\maketitle

There has been a strong interest in reaction studies involving so-called halo nuclei since their anomalously large interaction radii were discovered more than 25 years ago \cite{Tanihata}. The definition of a halo nucleus is still being debated, but at least three conditions are required \cite{Alkhalili}: (i) low separation energy of the valence particle (or particle cluster); (ii) a wave function that is in a low relative angular momentum state (preferably an s-wave); (iii) decoupling from the core. The nucleus $^{11}$Li, a two-neutron halo nucleus, is a prime example.\\

Possible candidates for one-neutron halo nuclei are still being debated. For example, in Ref. \cite{Alkhalili} $^{11}$Be, $^{15}$C, $^{19}$C and $^{23}$O are listed as candidates, based mainly on the width of their momentum distributions measured in one-neutron removal reactions \cite{Sauvan}. The latter two isotopes are located far away from the valley of stability and are currently not available with beam intensities sufficient to allow for studies of fusion reactions. The remaining two nuclei, $^{11}$Be and $^{15}$C, are closer to the valley of $\beta$ stability and can be produced with higher beam intensities. Measurements of interaction radii for $^{11}$Be and $^{15}$C at a high bombarding energy of 950 MeV/u \cite{Ozawa1} showed a radius increase only for $^{11}$Be. Later studies at lower energies (E$\sim$83 and 51 MeV/u) \cite{Fang}, however, demonstrated an increase in the interaction radius of $^{15}$C when compared to those of the neighboring $^{14,16}$C isotopes. \\

The ground state of $^{15}$C can be described as an $s_{1/2}$ neutron coupled to a $^{14}$C core with a separation energy of 1.218 MeV and a spectroscopic factor of $\sim$1 as measured in the $^{14}$C(d,p)$^{15}$C reaction \cite{Murillo}. In comparison, $^{11}$Be has a smaller neutron separation energy (0.503 MeV), but a slightly smaller one-neutron spectroscopic factor (see Ref. \cite{Aumann} and references quoted therein). Several fusion experiments with $^{11}$Be beams have been performed \cite{Signorini1,Signorini3,Signorini4}. Their interpretation suffers from the difficulty of not having a good spherical nucleus as a reference system. Early experiments used $^9$Be \cite{Signorini4}, which also has a low neutron binding energy of 1.665 MeV. Later studies replaced $^9$Be with the even-even nucleus $^{10}$Be, yet no enhancement in the fusion cross sections was found in either case. For the one-proton halo nucleus $^{17}$F no fusion enhancement was observed experimentally \cite{Rehm1}. This observation was explained through the polarization of the incoming $^{17}$F projectile in the Coulomb field of the $^{208}$Pb target nucleus, which keeps the weakly-bound proton away from the interaction zone. However, there is at present no consensus on the behavior of the low-energy fusion cross sections induced by halo nuclei. Reviews on this topic have been published in Refs. \cite{liang,loveland,canto,keeley}, and several theoretical predictions can be found in the literature \cite{Signorini2,theo1,theo2,theo3,theo4}. These calculations include coupling to soft dipole modes and to transfer and breakup channels. Both enhancement and suppression of the fusion cross sections at low energies has been predicted. \\

In this letter we report on a measurement of fusion in the $^{15}$C + $^{232}$Th system, studied by detecting the fission fragments produced from the decay of the excited $^{247}$Cm compound nucleus. Two coincident fission fragments emitted with high energies can be detected with good efficiency and provide a clean signal for the fusion-fission process. Furthermore, when compared to $^{238}$U, $^{232}$Th has the advantage that transfer-induced fission is expected to have a much smaller cross section, as discussed below. Since $^{15}$C is located next to $^{14}$C, which has a closed neutron shell \mbox{(S$_n$=8.177 MeV)}, a measurement of fusion-fission cross sections for $^{14}$C + $^{232}$Th provides a good reference reaction involving a spherical projectile. \\

The measurements of the fusion excitation functions were done in three steps. First, the excitation function for the system $^{13}$C+$^{232}$Th was measured and used to determine the detection efficiency by normalizing the data to the results from Ref. \cite{Kumar}. Then, an excitation function for the system $^{14}$C+$^{232}$Th was measured, providing a reference involving a closed shell nucleus. This was then followed by the measurement involving the halo nucleus $^{15}$C. \\

The experimental setup for the fusion-fission experiment was similar to the one described in Ref. \cite{Rehm1}. Four \mbox{5x5 cm$^2$} Si surface barrier detectors subdivided into four quadrants surrounded the $^{232}$Th target \mbox{(640 $\mu$g/cm$^2$)} with two pairs opposite to each other, covering the angular range between $25^\circ-70^\circ$ and $115^\circ-160^\circ$. Fusion-fission events were identified by the detection of coincident high-energy particles in two opposite detectors. At a distance of \mbox{6 cm}, the four detectors provided an average detection efficiency of 5.1\% as calculated by a Monte Carlo simulation. In order to be independent of the calculated efficiency, which depends on the angle between the two fission fragments, the measured fusion-fission yields from $^{13}$C + $^{232}$Th were normalized to the data from Ref. \cite{Kumar}. The detection efficiency was determined to be 5.3\%, in good agreement with the Monte Carlo simulation.\\

To expedite beam energy changes for the fusion excitation functions, Au foils with thicknesses between 4.9 and \mbox{14.9 mg/cm$^2$} were inserted into the beam \mbox{$\sim$55 cm} upstream from the target. The degrader foils reduced the beam intensities by factors of 3-5. The energy loss of the ions in the $^{232}$Th target and in the degrader foils, and the stability and purity of the beams during the runs, were monitored by detecting elastically scattered beam particles at $\theta_{lab}$=4.8$^\circ$ in the Enge split-pole spectrograph which was located behind the Si-detector array. The energy width of the $^{15}$C beam with e.g. a 14.9 mg/cm$^2$ Au foil was measured to be $\sim$720 keV (FWHM). Since the count rate at the small scattering angle of the spectrograph is very sensitive to the beam profile, the relative normalization of the $^{13,14,15}$C measurements was achieved by using the elastically scattered particles detected in the four most forward quadrants of the Si detectors ($\theta \sim35^\circ$) located symmetrically around of the beam.\\

\begin{figure}
\includegraphics[width=1.\columnwidth]{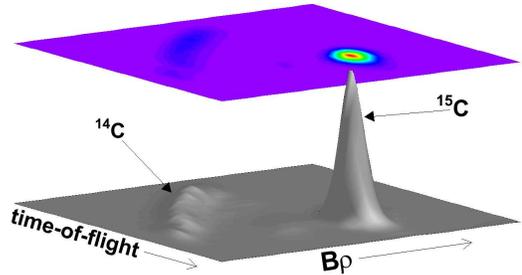}
\caption{\label{fig1} Color online: Spectrum of time-of-flight vs. magnetic rigidity B$\rho$ for a mixed radioactive $^{14,15}$C beam scattered elastically off a $^{232}$Th target and detected in the focal plane of the magnetic spectrograph. The beam passed through a 13.2 mg/cm$^2$ thick Au foil in order to reduce its energy from 73.95 MeV to 59.80 MeV.}  
\end{figure}

The $^{15}$C beam was produced via the In-Flight Technique \cite{Harss} by bombarding a cryogenically cooled gas cell filled with deuterium at 1.4 atm with an intense ($\sim$100 pnA) $^{14}$C beam delivered by the ATLAS accelerator. The $^{15}$C ions produced via the d($^{14}$C,$^{15}$C)p reaction were focused with a superconducting solenoid located behind the gas cell and rebunched with a superconducting resonator. The beam intensity of the $^{15}$C beam was in the range of \mbox{1$\times 10^6$ s$^{-1}$} (E=73.95 MeV) to \mbox{2.5$\times 10^5$ s$^{-1}$} (E=57.51 MeV). The main contaminants in the $^{15}$C beam were $^{14}$C ions scattered in the production target. An RF sweeper system \cite{pardo} located midway between the production target and the experimental setup reduced the $^{14}$C contamination to 3-28\% for the different energies. Fig. \ref{fig1} gives a spectrum of time-of-flight vs. magnetic rigidity (B$\rho$) measured in the focal plane of the spectograph for the $^{15}$C beam attenuated with a 13.2 mg/cm$^2$ Au foil to an energy of 59.8 MeV. The main peak containing $\sim$72\% of the total yield originates from $^{15}$C ions, while the group at lower magnetic rigidities results from scattering of the primary $^{14}$C beam. Particle identification was obtained using the time-of-flight and the $\Delta$E signals from the focal plane detector. In the previous $^{17}$F study \cite{Rehm1}, the remnants from the primary $^{17}$O beam did not contribute to the fusion measurements since the energy of the contaminant particles was lower by the square of the ratio of the charge states (64/81). In this experiment, the contaminant $^{14}$C$^{6+}$ particles have energies that can be higher than those of $^{15}$C$^{6+}$. Thus, the cross sections for the $^{15}$C + $^{232}$Th reaction need to be corrected for the contributions from the $^{14}$C + $^{232}$Th reaction. For this correction, the B$\rho$ spectrum for $^{14}$C (see Fig. \ref{fig1}) was converted into an energy spectrum, the yield corrected for the E$^{-2}$ dependence of the Rutherford cross section, and then folded with the fusion-fission cross section for $^{14}$C + $^{232}$Th measured earlier. These corrections to the cross sections are negligible at the two highest energies and increase to 13\% for the lowest energy point. \\

\begin{figure}
\includegraphics[width=1.\columnwidth]{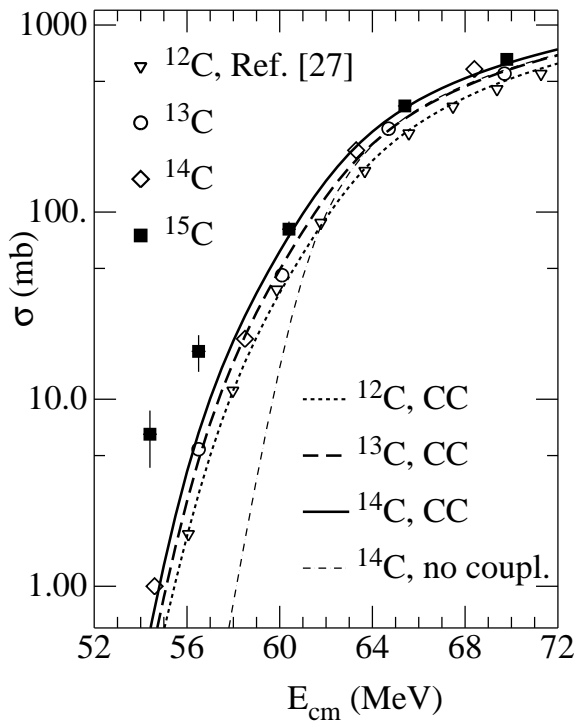}
\caption{\label{fig2} Cross section of the fusion-fission reactions $^{13,14,15}$C + $^{232}$Th vs. c.m. energy for the reactions studied in this experiment. If errors bars are not shown, the uncertainties are smaller than the symbols. Data for the system $^{12}$C + $^{232}$Th from Ref. \cite{transfer1} are also included. The lines are the result of coupled-channels calculations for $^{12,13,14}$C. The coupled-channel calculation for $^{15}$C overlaps with that of $^{14}$C and is therefore not included. See text for details. }  
\end{figure}

Fig. \ref{fig2} provides the experimental fusion-fission cross sections induced by $^{13,14,15}$C ions on $^{232}$Th as a function of the center-of-mass energy. The energy spread of the beam measured in the spectograph is smaller than the width of the symbol. Also included are the cross sections for the system $^{12}$C + $^{232}$Th taken from Ref. \cite{transfer1}. The cross sections for $^{12,13,14}$C agree with the CC calculations within their experimental uncertainties, while the fusion-fission cross section for $^{15}$C + $^{232}$Th is enhanced at the lowest energies by factors of $\sim$5 with respect to the CC calculation and $^{13,14}$C experimental data. This enhancement is similar to that seen in the fusion measurements with the four-neutron halo nucleus $^8$He on $^{197}$Au \cite{Lemasson}. The cross sections for the three systems are summarized in \mbox{Table \ref{tab1}}.\\

\begin{table}
\caption{\label{tab1}Fusion-fission cross sections for $^{13,14,15}$C + $^{232}$Th.}
\begin{ruledtabular}
\begin{tabular}{clclc|}
System & E$_{cm}$ (MeV) & $\sigma$ (mb) \\
\hline
  \\
 $^{13}$C + $^{232}$Th & 69.7(1) & 549(13)\\
  & 64.7(2) & 279(9)\\
  & 60.1(2) & 46(4)\\
  & 56.5(2) & 5.4(5)\\
  \\
 $^{14}$C + $^{232}$Th & 68.4(2) & 581(8)\\
   & 63.3(2) & 214(6)\\
  & 58.5(2) & 21(1)\\
  & 54.6(2) & 1.0(1)\\
  \\
 $^{15}$C + $^{232}$Th & 69.8(2) & 655(21)\\
   & 65.4(3) & 369(22)\\
  & 60.4(3) & 81(8)\\
  & 56.5(3) & 18(4)\\
  & 54.4(3) & 6.5(2.2)\\
  \end{tabular}
\end{ruledtabular}
\end{table}

The segmentation of the detectors used in these measurements did not permit a separation of transfer-induced fission reactions from fusion-fission. For nuclei close to $^{232}$Th, which can be populated in few-nucleon transfer reactions, the fission barriers are typically around 6 MeV \cite{Back}; i.e., comparable to the fission barrier of the compound nuclei $^{245,246,247}$Cm \cite{Maslov}. The excitation energies of the nuclei produced in transfer and compound-nucleus fusion reactions are, however, quite different (a few MeV for transfer reactions and $\sim$40 MeV  for fusion-fission). Thus, the contribution of transfer-induced fission is expected to be small. From experiments with stable beams in the immediate vicinity of $^{15}$C, the maximum contribution from transfer-induced fission of $^{11}$B, $^{12}$C, and $^{13}$C on $^{232}$Th, was found to be of the order of a few percent relative to the total fusion cross section \cite{Nadkarni,transfer1,Kumar}.\\

The situation is quite different for the fusion of $^6$He on $^{238}$U \cite{Trotta,Raabe}. There, a large fusion enhancement was reported in Ref. \cite{Trotta}. A later experiment, however, suggested that this enhancement originated from transfer-induced fission caused by the two-neutron transfer reaction ($^6$He,$^4$He) \cite{Raabe}. The Q-value for the $^{238}$U($^6$He,$^4$He) reaction is 9.76 MeV, which means that, due to Q-matching conditions, the main strength of the transfer yield is at excitation energies of $\sim$10 MeV, well above the fission barrier in $^{240}$U where the fission probability reaches values of 30\% \cite{Back1}. Thus, a large fraction of the transfer yield results in fission of the residual nuclei in this case.\\

The Q-value for the one-neutron transfer reaction $^{232}$Th($^{15}$C,$^{14}$C) is 3.57 MeV. No cross section measurements for this reaction exist, so we have used the systematics of neutron transfer obtained from reactions with stable beams in nearby systems \cite{Rehm2,arns}. The location and the width of the Q-window was calculated with the DWBA program PTOLEMY \cite{PTOLEMY}. In $^{233}$Th, only states at excitation energies above 6 MeV can contribute to transfer-induced fission. The total cross section for the ($^{15}$C,$^{14}$C) reaction is estimated from systematics to be around 300 mb. Folding the Q-window with the energy-dependent fission probability for $^{233}$Th \cite{Back}, we obtain an upper limit for the transfer-fission yield of about 0.5 mb in the energy range of 54-60 MeV. This is smaller than the fission yields measured in this experiment by a factor of at least 10. Thus, similar to the results obtained for $^{11}$B, $^{12}$C, $^{13}$C and $^{16}$O on $^{232}$Th, transfer-induced fission is expected to be small for the $^{15}$C + $^{232}$Th system in the energy range measured in this experiment.  \\

To compare the $^{13,14,15}$C + $^{232}$Th fusion cross sections to the predictions of a coupled-channel treatment, calculations have been performed using the code from Ref. \cite{Esbensen}. These calculations use a deformed Woods-Saxon potential, with the target described by static quadrupole and hexadecapole deformations. Couplings to quadrupole and octupole excitations in the different carbon isotopes and to the 3$^-$ excitation in $^{232}$Th have been included. The results of these calculations for the systems $^{12,13,14}$C + $^{232}$Th are given in Fig. \ref{fig2} by the dotted ($^{12}$C), dashed ($^{13}$C), and solid ($^{14}$C) lines. The dot-dashed curve is the result of a standard, no-coupling barrier-penetration calculation which underpredicts the data by many orders of magnitude. To describe the fusion of $^{15}$C with $^{232}$Th it was assumed that the valence neutron can be treated as a spectator. Thus, we assume that $^{15}$C has the same excitation spectrum as $^{14}$C, and that the ion-ion potential is the same as the one used in the calculations for $^{14}$C+$^{232}$Th. The valence neutron only appears in the calculations through the mass of $^{15}$C. The fusion cross sections predicted in this \textquoteleft spectator model' for $^{15}$C (not shown in Fig. \ref{fig2}) are essentially identical to the calculated cross sections for $^{14}$C+$^{232}$Th. A comparison to the experimental data indicates that, at high energies, the fusion data for $^{15}$C+$^{232}$Th are consistent with the spectator model. Below E$_{cm}\sim$59 MeV, additional effects come into play which lead to a strongly enhanced cross section. \\        

In summary, the longstanding question of whether the fusion of nuclei involving weakly bound particles is enhanced or suppressed at low energies has been addressed for the system $^{15}$C + $^{232}$Th. We find that the fusion-fission cross section is enhanced by a factor of 5 in comparison to those for $^{12,13,14}$C at the lowest energies studied in this experiment. This enhancement is at variance with the calculations of Refs. \cite{theo3,theo4}. In Ref. \cite{theo3} a reduction of the fusion cross section at these energies was predicted, while in Ref. \cite{theo4} there is little effect from transfer reactions on the fusion cross sections at the energies of interest. Using an improved detection setup and higher intensity beams, an extension of these measurements towards lower cross sections by another order of magnitude is feasible. A measurement of the transfer channels in $^{15}$C+$^{232}$Th would also be of interest, though such a measurement is currently beyond the limits of what is possible with existing capabilities. \\    

We would like to thank D. Hinde (ANU) for providing the $^{12}$C + $^{232}$Th data in tabulated form, and K. L. Jensen (Aarhus University, Denmark) for valuable discussions. We also want to thank the ATLAS operations staff for producing the carbon beams. This work was supported by the US Department of Energy, Office of Nuclear Physics, under contract DE-AC02-06CH11357 (ANL) and DE-FG02-04ER41320 (WMU) and by the NSF JINA grant PHY0822648.

\end{document}